%% file: iclr2024_conference.tex
\documentclass{article}

\usepackage{arxiv}

\usepackage[utf8]{inputenc} 
\usepackage[T1]{fontenc}    
\usepackage{hyperref}       
\usepackage{url}            
\usepackage{booktabs}       
\usepackage{nicefrac}       
\usepackage{microtype}      
\usepackage{graphicx}
\usepackage{natbib}
\usepackage{doi}
\usepackage{color}
\usepackage{colortbl}
\usepackage{multicol}
\usepackage{amsmath,amsfonts,amssymb,stmaryrd,xspace,enumitem}
\usepackage{multirow}
\usepackage{algorithmic,algorithm}
\usepackage{subfigure}
\usepackage{verbatim}
\usepackage{wrapfig}
\usepackage{tikz}
\usetikzlibrary{shapes.geometric, positioning}

\input{math_commands}

\title{\puma: Secure Inference of LLaMA-7B in Five Minutes}

\input{authors}

\begin{document}

\maketitle
\begin{abstract}
With ChatGPT as a representative, tons of companies have began to provide services based on large Transformers models. However, using such a service inevitably leak users' prompts to the model provider. Previous studies have studied secure inference for Transformer models using secure multiparty computation (MPC), where model parameters and clients' prompts are kept secret. Despite this, these frameworks are still limited in terms of model performance, efficiency, and deployment. To address these limitations, we propose framework \puma\ to enable fast and secure Transformer model inference. 
Our framework designs high quality approximations for expensive functions such as $\gelu$ and $\softmax$, and significantly reduce the cost of secure inference while preserving the model performance. Additionally, we design secure Embedding and LayerNorm procedures that faithfully implement the desired functionality without undermining the Transformer architecture. \puma\ is about $2\times$ faster than the state-of-the-art  framework \mpcformer (ICLR 2023) and has similar accuracy as plaintext models without fine-tuning (which the previous works failed to achieve).  
\puma\ can even evaluate LLaMA-7B in around 5 minutes to generate $1$ token. To our best knowledge, this is the first time that a model with such a parameter size is able to be evaluated under MPC. \puma\ has been open-sourced in the Github repository of SecretFlow-SPU\footnote{\url{https://github.com/secretflow/spu/tree/main/examples/python/ml/flax_llama7b}}.

\end{abstract}

\input{Introduction}
\input{relatedwork}
\input{prem}

\input{overview}
\input{experiments}
\input{conclusion}

\bibliography{iclr2024_conference}
\bibliographystyle{iclr2024_conference}

\input{appendix}

\end{document}

%% file: math_commands.tex
\usepackage{amsmath,amsfonts,bm}

\def\eqref#1{equation~\ref{#1}}

\def\1{\bm{1}}

\DeclareMathAlphabet{\mathsfit}{\encodingdefault}{\sfdefault}{m}{sl}
\SetMathAlphabet{\mathsfit}{bold}{\encodingdefault}{\sfdefault}{bx}{n}

\newcommand{\E}{\mathbb{E}}

\newcommand{\softmax}{\mathrm{softmax}}

\def\ie{\textit{i.e.}} 

\def\eg{\textit{e.g.}}

\def\puma{\textsc{Puma}}
\def\mpcformer{\textsc{MPCFormer}}
\definecolor{mygray}{gray}{0.95}

\newcommand{\gelu}{\mathsf{GeLU}}

\newcommand{\layernorm}{\mathsf{LayerNorm}}
\newcommand{\ffn}{\mathsf{FFN}}
\newcommand{\attention}{\mathsf{Attention}}
\newcommand{\Q}{\mathbf{Q}}
\newcommand{\K}{\mathbf{K}}
\newcommand{\V}{\mathbf{V}}
\newcommand{\M}{\mathbf{M}}
\newcommand{\x}{\mathbf{x}}
\newcommand{\y}{\mathbf{y}}
\newcommand{\z}{\mathbf{z}}

\newcommand{\ZL}{{\mathbb{Z}_{2^\ell}}}

\newcommand{\tokenid}{\mathsf{id}}
\newcommand{\e}{\mathbf{e}}

\newcommand{\share}[1]{\ensuremath{\llbracket #1\rrbracket}\xspace}

\newcommand{\shareb}[1]{\ensuremath{\llbracket #1\rrbracket^\mathsf{B}}\xspace}


%% file: authors.tex
\author{{Ye Dong}
\\
Ant Group \\
\texttt{dongye.dong@antgroup.com} \\
\And
{Wen-jie Lu} \\
Ant Group \\
\texttt{juhou.lwj@antgroup.com}\\ 
\And
{Yancheng Zheng} \\
Ant Group\\
\texttt{zhengyancheng.zyc@antgroup.com}\\
\And
{Haoqi Wu}\\
Ant Group\\
\texttt{haoqi.whq@antgroup.com}\\
\And
{Derun Zhao}\\
Ant Group\\
\texttt{zhaoderun.zdr@antgroup.com}\\
\And
{Jin Tan}\\
Ant Group\\
\texttt{tanjin.tj@antgroup.com}\\
\And
{Zhicong Huang}\\
Ant Group\\
\texttt{zhicong.hzc@antgroup.com}\\
\And
{Cheng Hong}\\
Ant Group\\
\texttt{vince.hc@antgroup.com}\\
\And
{Tao Wei}\\
Ant Group\\
\texttt{lenx.wei@antgroup.com}\\
\And
{Wenguang Chen}\\
Ant Group\\
\texttt{yuanben.cwg@antgroup.com}
}

%% file: Introduction.tex
\section{Introduction}\label{sec:intro}
Pre-trained Transformer models~\citep{transformer} have attracted much attentions for their high performance in practical tasks~\citep{gpt,zhuge2021kaleido} and been widely in Deep Learning as a Service (DLaaS) paradigm~\citep{soifer2019deep}. However, these services can raise privacy concerns, such as in the case of ChatGPT~\citep{chatgpt}, which requires either users to reveal their private prompts to the service provider or the service provider to release their proprietary trained weights to users.

One solution to address the privacy concerns of Transformer models service is Secure Multi-Party Computation (MPC)~\citep{shamir1979share,yao1986generate,gmw}, which can keep data and model weights private during inference. 
\citep{hao2022iron,li2023mpcformer,privformer,liang2023merge,liu2023llms} have proposed various ways to support  secure Transformer models inference, but these approaches still have one or several of the following drawbacks: 

 \textbf{High inference cost.} Non-linear functions like $\gelu$ and $\softmax$ are challenge to design in MPC. \citep{hao2022iron} computes these non-linear functions in a faithful way. \eg, they design $\gelu$ using $\tanh$ based on general MPC exponentiation method proposed by~\citep{rathee2021sirnn}. But these general methods are quite expensive in terms of computation and communication, and only tested under small bitwidth (e.g. below 32). 

\textbf{Retraining required.} 
To reduce the cost of non-linear functions, several works \citep{li2023mpcformer,privformer,liu2023llms} suggested to approximate $\gelu$ and $\softmax$ using simpler functions like ReLU and quadratics. These functions are  up to an order of magnitude cheaper in MPC, but would introduce utility loss to the Transformer model. As a result, they require an extra step of model retraining (fine-tuning).  However, retraining is unfriendly for data-limited participants, and  might not achieve satisfactory performance~\citep{kumar2022fine}. 

 \textbf{Incompatible architectures.} 
\citep{li2023mpcformer,liang2023merge} proposed to modify the architecture of Transformer models to further accelerate secure inference, \eg, decompose the embedding procedure or reorganize the linear layers. 
Worsely, \citep{li2023mpcformer} does not support secure LayerNorm and  simulated the costs using BatchNorm, resulting in incorrect secure inference results.
These modifications are in conflicts with existing plaintext Transformer systems, and would lead to deployment obstacles. 


To summarize, in the field of MPC Transformer inference, achieving both model performance and efficiency is challenging, and people may ask the following question:

\textit{Could pre-trained large transformer models be securely and efficiently evaluated with similar accuracy as in plaintext, without further retraining ?}

 To address this challenge, we propose the \puma\ framework, which is a fast and accurate end-to-end secure Transformer inference framework. Our contributions can be summarized as follows:
\begin{itemize}
    \item \textbf{New  Approximations for Non-linear Functions.} We propose more accurate and faster approximations for the expensive non-linear functions (\eg, $\gelu$ and $\softmax$) in Transformer models. Different from existing works, we design the approximations based on the specialized properties of these non-linear functions to achieve both accuracy and efficiency. 

    \item \textbf{Faster and More Accurate  Secure Inference.}   We make extensive experiments on 6 transformer models and 4 datasets,  the results show that \puma's precision is similar to plaintext ones'  and is about $2\times$ faster than \mpcformer\  (note that \mpcformer\ does not achieve similar precision as \puma). \puma\ can even evaluate LLaMA-7B in around 5 minutes to generate one word. To our best knowledge, this is the first time that such a large language model is able to be evaluated under MPC.
        
    \item \textbf{End-to-End Framework compatible with plaintext.}
    We design and implement all the layers required by Transformer (including the Embedding and LayerNorm layers that are missing in other works)  in MPC.  This allows us to load and securely evaluate the pre-trained plaintext Transfomer models (\eg\ downloaded from Hugging face) easily. To our best knowledge,  \puma\ is the first open-sourced MPC solution  that supports accurate inference of pre-trained Transformer models without further modifications such as re-training.

\end{itemize}

\textbf{Organization.} We summarize the related work in \S~\ref{sec:relatedwork} and present the background in \S~\ref{sec:back}. We give \puma's high-level view and concrete design in \S~\ref{sec:design}. We analyze the experimental results in \S~\ref{sec:experiment} and conclude this work in \S~\ref{sec:conclusion}.

%% file: relatedwork.tex
\section{Related Work}\label{sec:relatedwork}

Secure Multiparty Computation (MPC)~\citep{yao1986generate,gmw} enables distrusted parties to jointly compute a function while keeping their  inputs private, and secure deep learning inference using MPC has gained much attention due its high privacy protection.
These works operate in a variety of  models and architectures, including two-party setting~\citep{mohassel2017secureml,liu2017oblivious,mishra2020delphi,cheetah,patra2021aby2,cryptflow2}, three-party setting~\citep{wagh2019securenn,aby3,wagh2020falcon,kumar2019cryptflow,patra2020blaze,tan2021cryptgpu,meteor}, or four-party setting~\citep{byali2020flash,dalskov2021fantastic}. 
However, most of these approaches only consider secure inference of convolutional/deep neural networks, and cannot be directly extended to support Transformer models. Recently several research works \citep{hao2022iron,li2023mpcformer,privformer,liang2023merge,liu2023llms} have proposed MPC-based secure inference solutions for Transformer models, but these approaches still have limitations in terms of model performance, efficiency, and deployment. Among these works, \mpcformer~\citep{li2023mpcformer} is the only one that have been open-sourced, it is based on CrypTen~\citep{crypten2020} which is a  three-party framework that uses a non-colluding third party to produce correlated randomness for the client and server. Also their  three-party model with non-colluding assumption has the highest concrete efficiency among different MPC settings. So we mainly compare our proposed framework \puma\ with \mpcformer\ under the same three-party setting.

%% file: prem.tex
\section{Background}\label{sec:back}

\subsection{Notations}\label{sec:notations}
The main used notations are as follows: $P_i$ represents the $i$-th computing party, $i \in \{0,1,2\}$. The uppercase bold letter $\mathbf{X}$ is used for matrices, and the lowercase bold letter $\x$ denotes vectors. $\x[i]$ denotes the $i$-th element of vector $\x$, while lowercase letter $x$ is used for scalar values. $\ZL$ denotes the discrete ring modulo $2^\ell$, $\mathbb{R}$ denotes real numbers.
$\share{\cdot}$ is used for 2-out-of-3 replicated secret sharing~\citep{araki2016high,aby3}.

\subsection{Transformer Model}\label{sec:transformer}
Transformer models have achieved remarkable success in language understanding~\citep{gpt,bert,xlnet,touvron2023llama}, vision understanding~\citep{zhuge2021kaleido,dong2022bootstrapped,chen2021pre}, and etc. Two popular variants are Bert (Bidirectional Encoder Representations from Transformers)~\citep{bert} and GPT (Generative Pre-Trained models)~\citep{gpt}. A Transformer model~\citep{transformer} mainly consists of Embedding,  \textbf{Attention}, \textbf{Feed-Forward Network}, and \textbf{LayerNorm} sub-layers:

\textbf{Attention.} Given inputs $(\Q, \K, \V)$, the $\attention$ function is computed as 
$\attention(\Q,\K,\V)=
\softmax(\Q \cdot \K^\mathsf{T} + \M) \cdot \V$,
where $\M$ can be viewed as a bias matrix. Besides, \citep{transformer} proposed Multi-Head $\attention$ to jointly attend to information from different representation subspaces at different positions.

\textbf{Feed-Forward Network ($\ffn$).} 
$\ffn$ is applied to each position separately and identically. This consists of two linear transformations with an activation in between, and the most commonly used activation function is $\gelu$. Given input $\x$ and parameters $\{\mathbf{W}_1, \mathbf{b}_1, \mathbf{W}_2, \mathbf{b}_2\}$, $\ffn$ can be formalized as $\ffn(\x) = \mathbf{W}_2\gelu(\mathbf{W}_1\x+\mathbf{b}_1)+\mathbf{b}_2$.
Note that the parameters of linear transformations are different from layer to layer.

\textbf{LayerNorm.} Given vector $\x\in \mathbb{R}^n$, $\layernorm$ is defined as: $\layernorm(\x)[i] =  \gamma \cdot \frac{\x[i]-\mu}{\sqrt{\sigma}} + \beta$, where $(\gamma, \beta)$ are trained parameters, $\mu = \frac{\sum_{i=1}^n \x[i]}{n}$, and $\sigma = \sum_{i=1}^n (\x[i] - \mu)^2$.


\subsection{2-out-of-3 Replicated Secret Sharing}\label{sec:3partyss}
A secret value $x\in \ZL$ is shared by three random values $x_0, x_1, x_2 \in \ZL$ with $x=x_0+x_1+x_2 \pmod{2^\ell}$. In 2-out-of-3 replicated secret sharing (denoted as $\share{\cdot}$-sharing), party $P_i$ gets $\share{x}_i = (x_i, x_{i+1})$. Without special declaration, we compute in $\ZL$ and omit $\pmod{2^\ell}$ for brevity.
In the case of $\ell>1$ (\eg, $\ell=64$) which support arithmetic operations (\eg, $+$, $-$, and $\cdot$), we refer to this type as \textit{Arithmetic Sharing} and use notation $\share{\cdot}$. \textit{Boolean Sharing} ($\shareb{\cdot}$) refers to $\ell=1$ where $(+,-)$ and $\cdot$ are respectively replaced by bit-wise $\oplus$ and $\land$.

\textbf{Addition.}
Let $(c_1$, $c_2$, $c_3)$ be public constants, and $(\share{x}, \share{y})$ be two secret-shared values. Then, $\share{c_1 x+c_2 y+c_3}$ can be computed as $(c_1 x_0+c_2 y_0+c_3, c_1 x_1+c_2 y_1, c_1 x_2+c_2 y_2)$ where $P_i$ can compute its share locally. When $(c_1 =1,c_2 =1,c_3 =0)$, we get $\share{x+y}$.  

\textbf{Multiplication.}
In secure multiplication protocol $\Pi_{\mathsf{Mul}}$, given two shared values $\share{x}$ and $\share{y}$, parties follows steps: \romannumeral1) First, $P_i$ computes $z_i = x_iy_i + x_{i+1}y_i + x_iy_{i+1}$ locally, \romannumeral2) Parties then perform \textit{re-sharing} by letting $P_i$ sends $z_i'=\alpha_i+z_i$ to $P_{i-1}$, where $\alpha_0+\alpha_1+\alpha_2=0$ ($P_i$ can generate $\alpha_i$ in the setup phase as \cite{aby3}). \romannumeral3) Finally, $\{(z_0', z_1'), (z_1', z_2'), (z_2',z_0')\}$ form $\share{x\cdot y}$. 

\textbf{Underlying Protocols.}
In addition to addition and multiplication, \puma\ relies on several other underlying protocols: boolean-arithmetic multiplication ($\Pi_{\mathsf{Mul_{BA}}}$), square $\Pi_{\mathsf{Square}}$, equality test ($\Pi_{\mathsf{Eq}}$), less than ($\Pi_{\mathsf{LT}}$), reciprocal ($\Pi_{\mathsf{Recip}}$), maximum ($\Pi_{\mathsf{Max}}$), and reciprocal of square root ($\Pi_{\mathsf{rSqrt}}$), from the state-of-the-art works. We employ them in a black-box manner, and only enumerate the inputs and outputs of these protocols as follows: 
\begin{itemize}
\begin{multicols}{2}
    \item $\share{z}=\Pi_{\mathsf{Mul_{BA}}}(\shareb{b}, \share{x})$, s.t. $z=b\cdot x$
    \item $\share{z}=\Pi_{\mathsf{Square}}(\share{x})$, s.t. $z=x^2$
    \item $\shareb{z}=\Pi_{\mathsf{Eq}}(\share{x},\share{y})$, s.t. $z=1\{x=y\}$
    \item $\shareb{z}=\Pi_{\mathsf{LT}}(\share{x}, \share{y})$, s.t. $z=1\{x<y\}$
    \item $\share{z}=\Pi_{\mathsf{Recip}}(\share{x})$, s.t. $z=1/x$
    \item $\share{z}=\Pi_{\mathsf{rSqrt}}(\share{x})$, s.t. $z=1/\sqrt{x}$
    \item $\share{z}=\Pi_{\mathsf{Max}}(\share{\x})$, s.t. $z=\mathsf{maximum}(\x)$
\end{multicols}
\end{itemize}
$1\{e\}$ returns $1$ that when condition $e$ is \textsf{true}, and $0$ otherwise. 
For detailed protocol constructions, please refer to~\citep{aby3,rSqrt,keller2020mp}.

\textbf{Fixed-Point Representation \& Truncation.}
Real numbers has to be encoded into fixed-point numbers before represented in finite rings/fields. To avoid overflow,  $\Pi_{\mathsf{Trunc}}^{f}$ has to be used after each fixed-point multiplication to truncate the least $f$ bits securely. For simpler description, we include $\Pi_{\mathsf{Trunc}}^{f}$ in $\Pi_{\mathsf{Mul}}$ and $\Pi_{\mathsf{Square}}$ by default and and do not explicitly mention it in our protocol designs.

The above operations can be easily extended to vectors and matrices, and we use the same notation for vector and matrix operations for simplicity. For more details, please refer to~\citep{aby3,wagh2020falcon}.

\textbf{Threat Model.}
Following previous works~\citep{aby3,li2023mpcformer},
\puma\ is secure against a semi-honest adversary  that corrupts no more than one of the three computing parties. Semi-honest means such an adversary will follow the protocol specifications, but may try to learn other's private information during the protocol. Please note that \puma\ cannot defend against attacks based on inference results, and the mitigation of such attacks (\eg, differential privacy~\citep{abadi2016deep}) falls outside the scope of this study.

%% file: overview.tex
\section{Secure Design of \puma}\label{sec:design}
In this section, we first present an overview of \puma, and present the protocols for secure $\gelu$ , $\softmax$, embedding, and $\layernorm$ used by \puma. Note that the linear layers such as matrix multiplication are straightforward in replicated secret sharing, so we mainly describe our protocols for non-linear layers in this manuscript.

\subsection{Overview of \puma}\label{sec:overview}
To achieve secure inference of Transformer models, \puma\ defines three kinds of roles: one model owner, one client, and three computing parties. The model owner and the client  provide their models or inputs to the computing parties (i.e., $P_0$, $P_1$, and $P_2$) in a secret-shared form, then the computing parties execute the MPC protocols and send the results back to the client. Note that the model owner and client can also act as one of the computing party, we describe them separately for generality. \eg, when the model owner acts as $P_0$, the client acts as  $P_1$, a third-party dealer acts as $P_2$, the system model becomes the same with \mpcformer~\citep{li2023mpcformer}.

During the secure inference process, a key invariant is maintained: For any layer, the computing parties always start with 2-out-of-3 replicated secret shares of the previous layer's output and the model weights, and end with 2-out-of-3 replicated secret shares of this layer's output. As the shares do not leak any information to each party, this ensures that the layers can be sequentially combined for arbitrary depths to obtain a secure computation scheme for any Transformer-based model.

\input{Protocols/Gelu}

\input{Protocols/Softmax}
\input{Protocols/Embedding}

\input{Protocols/LayerNorm}

%% file: Protocols/Gelu.tex
\subsection{Protocol for Secure GeLU}\label{sec:gelu}
Most of the current approaches view the $\gelu$ function as a composition of smaller functions and try to optimize each piece of them, making them to miss the
chance of optimizing the private $\gelu$ as a whole. Given the $\gelu$ function:
\begin{equation}\label{eq:gelu}
\begin{split}
    \gelu(x) &= \frac{x}{2} \cdot \left(1 + \tanh \left( \sqrt{\frac{2}{\pi}} \cdot \left(x + 0.044715 \cdot x^3 \right) \right) \right)\\
    &\approx x\cdot \mathsf{sigmoid}(0.071355\cdot x^3 + 1.595769\cdot x) 
\end{split},
\end{equation}
these approaches~\citep{hao2022iron,characmpctranformer} focus either on designing approximate protocols for function $\tanh$
or using  existing general MPC protocols of exponentiation and reciprocal for $\mathsf{sigmoid}$. 

However, none of current approaches have utilized the fact that $\gelu$ function is almost linear on the two sides (\ie, $\gelu(x)\approx 0$ for $x<-4$ and $\gelu(x)\approx x$ for $x>3$). 
Within the short interval $[-4,3]$ of $\gelu$,
we suggest a piece-wise approximation of low-degree polynomials is a more efficient and easy-to-implement choice for its secure protocol. Concretely, our piece-wise low-degree polynomials are shown as equation~(\ref{eq:geluapprox}):
\begin{equation}\label{eq:geluapprox}
\gelu(x)=
\begin{cases}
0, & x<-4 \\
F_0(x), & -4 \le x < -1.95 \\
F_1(x), & -1.95 \le x \le 3 \\
x, & x >3
\end{cases},
\end{equation}
where polynomials $F_0()$ and $F_1()$ are computed by library $\mathsf{numpy.ployfit}$\footnote{\url{https://numpy.org/doc/stable/reference/generated/numpy.polyfit.html}} as equation~(\ref{eq:f0f1}). Surprsingly, the above simple poly fit works very well and our $\mathsf{max\ error}< 0.01403$, $\mathsf{median\ error}< 4.41e-05$, and $\mathsf{mean\ error}< 0.00168$.
\begin{equation}\label{eq:f0f1}
\begin{cases}
F_0(x) &= -0.011034134030615728 x^3 -0.11807612951181953 x^2 \\
&- 0.42226581151983866 x -0.5054031199708174\\
F_1(x) &= 0.0018067462606141187x^6 -0.037688200365904236 x^4 \\
&+ 0.3603292692789629x^2 + 0.5x + 0.008526321541038084
\end{cases}
\end{equation}

Formally, given secret input $\share{x}$, our secure $\gelu$ protocol $\Pi_{\gelu}$ is constructed as algorithm~\ref{protocol:gelu}. 

\input{Protocols/geluprotocol}

%% file: Protocols/geluprotocol.tex
\begin{algorithm}[tp]
\caption{Secure $\gelu$ Protocol $\Pi_{\mathsf{GeLU}}$}\label{protocol:gelu}
\begin{algorithmic}[1]
\REQUIRE
$P_i$ holds the 2-out-of-3 replicate secret share $\share{x}_i$ for $i\in \{0,1,2\}$ 
\ENSURE
$P_i$ gets the 2-out-of-3 replicate secret share $\share{y}_i$ for $i\in \{0,1,2\}$, where $y=\gelu(x)$.

\STATE $P_0$, $P_1$, and $P_2$ jointly compute
\begin{equation*}
\begin{split}
&\shareb{b_0} = \Pi_{\mathsf{LT}}(\share{x}, -4),~~~\vartriangleright b_0 = 1\{x<-4\}\\
&\shareb{b_1} = \Pi_{\mathsf{LT}}(\share{x}, -1.95),~~~\vartriangleright b_1 = 1\{x<-1.95\} \\
&\shareb{b_2} = \Pi_{\mathsf{LT}}(3, \share{x}),~~~~~~\vartriangleright b_2 = 1\{3<x\}
\end{split}
\end{equation*}
and compute 
$\shareb{z_0} = \shareb{b_0} \oplus \shareb{b_1}$,
$\shareb{z_1} = \shareb{b_1} \oplus \shareb{b_2} \oplus 1$, and $\shareb{z_2}=\shareb{b_2}$. Note that $z_0 = 1\{-4\le x < -1.95\}$, $z_1 = 1\{-1.95\le x\le 3\}$, and $z_2 = 1\{x>3\}$.

\STATE Jointly compute $\share{x^2} = \Pi_{\mathsf{Square}}(\share{x})$, $\share{x^3} = \Pi_{\mathsf{Mul}}(\share{x}, \share{x^2})$, $\share{x^4} = \Pi_{\mathsf{Square}}(\share{x^2})$, and $\share{x^6} = \Pi_{\mathsf{Square}}(\share{x^3})$.

\STATE Computing polynomials $\share{F_0(x)}$ and $\share{F_1(x)}$ based on $\{\share{x}, \share{x^2}, \share{x^3}, \share{x^4}, \share{x^6}\}$ as equation~(\ref{eq:geluapprox}) securely.

\RETURN$\share{y} = \Pi_{\mathsf{Mul_{BA}}}(\shareb{z_0}, \share{F_0(x)}) + \Pi_{\mathsf{Mul_{BA}}}(\shareb{z_1}, \share{F_1(x)})+\Pi_{\mathsf{Mul_{BA}}}(\shareb{z_2}, \share{x})$.

\end{algorithmic}
\end{algorithm}

%% file: Protocols/Softmax.tex
\subsection{Protocol for Secure Softmax}\label{sec:secureatten}

In the function $\attention(\Q,\K,\V)=
\softmax(\Q \cdot \K^\mathsf{T} + \M) \cdot \V$, the key challenge is computing function $\softmax$. For the sake of numerical stability, the $\softmax$ function is computed as
\begin{equation}\label{eq:softmax}
    \softmax(\x)[i]=\frac{\exp(\x[i] - \bar{x} - \epsilon)}{\sum_i \exp(\x[i] - \bar{x} - \epsilon)},
\end{equation}
where $\bar{x}$ is the maximum element of the input vector $\x$. 
For the normal plaintext softmax, $\epsilon=0$. For a two-dimension matrix, we apply equation~(\ref{eq:softmax}) to each of its row vector.

Formally, our detailed secure protocol  $\Pi_{\softmax}$ is illustrated in algorithm~\ref{protocol:softmax}, where we propose two optimizations:
\begin{itemize}
\item 
For the first optimization, we set $\epsilon$ in equation~\ref{eq:softmax} to a tiny and positive
value, e.g., $\epsilon =
10^{-6}$, so that the inputs to exponentiation
in equation~\ref{eq:softmax} are all negative. We exploit the negative operands
for acceleration. Particularly, we compute the exponentiation using the Taylor series~\citep{tan2021cryptgpu} with a simple clipping
\begin{equation}\label{eq:negexp}
\mathsf{negExp}(x) = \begin{cases}
    0, &x < T_{\exp} \\
    (1+\frac{x}{2^t})^{2^t}, &x\in [T_{\exp},0].
\end{cases}
\end{equation}
Indeed, we apply the less-than for the branch $x < T_{\exp}$
The division by $2^t$ can be achieved using
$\Pi_{\mathsf{Trunc}}^t$ since the input is already negative. Also, we can
compute the power-of-$2^t$ using $t$-step sequences of square function $\Pi_{\mathsf{square}}$ and $\Pi_{\mathsf{Trunc}}^f$. Suppose our MPC program uses
$18$-bit fixed-point precision. Then we set $T_{\exp}=-14$ given $\exp(-14) < 2^{-18}$, and empirically set $t = 5$.

\item 
Our second optimization is to reduce the number of divisions, which ultimately saves computation and communication costs.
To achieve this, for a vector $\x$ of size $n$, we have replaced the operation $\mathsf{Div}(\x, \mathsf{Broadcast}(y))$ with $\x \cdot  \mathsf{Broadcast}(\frac{1}{y})$, where $y=\sum_{i=1}^n\x[i]$. By making this replacement, we effectively reduce $n$ divisions to just one reciprocal operation and $n$ multiplications.
This optimization is particularly beneficial in the case of the $\softmax$ operation. The $\frac{1}{y}$ in the $\softmax$ operation is still large enough to maintain sufficient accuracy under fixed-point values. As a result, this optimization can significantly reduce the computational and communication costs while still providing accurate results.
\end{itemize}

\input{Protocols/softmaxprotocol}

%% file: Protocols/softmaxprotocol.tex
\begin{algorithm}[tp]
\caption{Secure $\softmax$ Protocol $\Pi_{\softmax}$}\label{protocol:softmax}
\begin{algorithmic}[1]
\REQUIRE
$P_i$ holds the 2-out-of-3 replicate secret share $\share{\x}_i$ for $i\in \{0,1,2\}$, and $\x$ is a vector of size $n$. 
\ENSURE
$P_i$ gets the 2-out-of-3 replicate secret share $\share{\y}_i$ for $i\in \{0,1,2\}$, where $\y=\softmax(\x)$.

\STATE $P_0$, $P_1$, and $P_2$ jointly compute
$\shareb{\mathbf{b}} = \Pi_{\mathsf{LT}}(T_{\exp}, \share{\x})$ and the maximum $\share{\bar{x}} = \Pi_{\mathsf{Max}}(\share{\x})$.

\STATE Parties locally computes $\share{\hat{\x}} = \share{\x} - \share{\bar{x}} - \epsilon$, and jointly compute $\share{\z_0} = 1+  \Pi_{\mathsf{Trunc}}^t(\share{\hat{\x}})$.

\FOR{$j=1,2,\dots, t$}
\STATE $\share{\z_j} = \Pi_{\mathsf{Square}}(\share{\z_{j-1}})$.
\ENDFOR

\STATE Parties locally compute $\share{z} = \sum_{i=1}^n \share{\z[i]}$ and jointly compute $\share{1/z} = \Pi_{\mathsf{Recip}}(\share{z})$.

\STATE Parties jointly compute $\share{\z / z} = \Pi_{\mathsf{Mul}}(\share{\z}, \share{1/z})$

\RETURN $\share{\y} = \Pi_{\mathsf{Mul}_{\mathsf{BA}}}( \shareb{\mathbf{b}}, \share{\z / z})$.

\end{algorithmic}
\end{algorithm}

%% file: Protocols/Embedding.tex
\subsection{Protocol for Secure Embedding}\label{sec:embed}

The current secure embedding procedure described in~\citep{li2023mpcformer} necessitates the client to  generate a one-hot vector using the token $\tokenid$ locally. This deviates from a plaintext Transformer workflow where the one-hot vector is generated inside the model. As a result, they have to carefully strip off the one-hot step from the pre-trained models, and add the step to the client side, which could be an obstacle for deployment.

To address this issue, we propose a secure embedding design as follows. Assuming that the token $\tokenid\in [n]$ and all embedding vectors are denoted by $\E= (\e_1^T, \e_2^T, \dots, \e_n^T)$, the embedding can be formulated as $\e_{\tokenid} = \mathbf{E}[\tokenid]$. Given $(\tokenid, \E)$ are in secret-shared fashion, our secure embedding protocol $\Pi_{\mathsf{Embed}}$ works as follows:
\begin{itemize}
    \item The computing parties securely compute the one-hot vector $\shareb{\mathbf{o}}$ after receiving $\share{\tokenid}$ from the client. Specifically, $\shareb{\mathbf{o}[i]}=\Pi_{\mathsf{Eq}}(i,\share{\tokenid})$ for $i\in [n]$.
    \item The parties can compute the embedded vector via $\share{\e_{\tokenid}} = \Pi_{\mathsf{Mul_{BA}}}(\share{\E}, \shareb{\mathbf{o}})$, where  does not require secure truncation.
\end{itemize}
In this way, our $\Pi_{\mathsf{Embed}}$ does not require explicit modification of the workflow of plaintext Transformer models, at the cost of more $\Pi_{\mathsf{Eq}}$ and $\Pi_{\mathsf{Mul_{BA}}}$ operations.

%% file: Protocols/LayerNorm.tex
\subsection{Protocol for Secure LayerNorm}\label{sec:seclayernorm}
Recall that given a vector $\x$ of size $n$, $\layernorm(\x)[i] =  \gamma \cdot \frac{\x[i]-\mu}{\sqrt{\sigma}} + \beta$, where $(\gamma, \beta)$ are trained parameters, $\mu = \frac{\sum_{i=1}^n \x[i]}{n}$, and $\sigma = \sum_{i=1}^n (\x[i] - \mu)^2$. In MPC, the key challenge is the evaluation of the divide-square-root $\frac{\x[i]-\mu}{\sqrt{\sigma}}$ formula. To securely evaluate this formula, CrypTen sequentially executes the MPC protocols of square-root, reciprocal, and multiplication. However, we observe that $\frac{\x[i]-\mu}{\sqrt{\sigma}}$ is equal to $(\x[i]-\mu)\cdot \sigma^{-1/2}$. And in the MPC side, the costs of computing the inverse-square-root $\sigma^{-1/2}$ is similar to that of the square-root operation~\citep{rSqrt}. Besides, inspired by the second optimization of \S~\ref{sec:secureatten}, we can first compute $\sigma^{-1/2}$ and then $\mathsf{Broadcast}(\sigma^{-1/2})$ to support fast and secure $\layernorm(\x)$. And our formal protocol $\Pi_{\layernorm}$ is shown in algorithm~\ref{protocol:layernorm}.

\input{Protocols/layernormprotocol}

%% file: Protocols/layernormprotocol.tex
\begin{algorithm}[tp]
\caption{Secure $\mathsf{LayerNorm}$ Protocol $\Pi_{\mathsf{LayerNorm}}$}\label{protocol:layernorm}
\begin{algorithmic}[1]
\REQUIRE
$P_i$ holds the 2-out-of-3 replicate secret share $\share{\x}_i$ for $i\in \{0,1,2\}$, and $\x$ is a vector of size $n$. 
\ENSURE
$P_i$ gets the 2-out-of-3 replicate secret share $\share{\y}_i$ for $i\in \{0,1,2\}$, where $\y=\mathsf{LayerNorm}(\x)$.

\STATE $P_0$, $P_1$, and $P_2$ compute $\share{\mu} = \frac{1}{n}\cdot \sum_{i=1}^n\share{\x[i]}$ and $\share{\sigma} = \sum_{i=1}^n \Pi_{\mathsf{Square}}(\share{\x} - \share{\mu})[i]$.

\STATE Parties jointly compute $\share{\sigma^{-1/2}} = \Pi_{\mathsf{rSqrt}}(\share{\sigma})$.

\STATE Parties jointly compute $\share{\mathbf{c}} = \Pi_{\mathsf{Mul}}((\share{\x} - \share{\mu}), \share{\sigma^{-1/2}})$

\RETURN $\share{\y} = \Pi_{\mathsf{Mul}}(\share{\gamma}, \share{\mathbf{c}}) + \share{\beta}$.

\end{algorithmic}
\end{algorithm}

%% file: experiments.tex
\section{Experimental Evaluations}\label{sec:experiment}

\textbf{Implementation.}
We implement \puma\ on top of SecretFlow-SPU~\citep{spu} in \textrm{C++} and Python. 
We encode the data in a fixed-point form under ring $\mathbb{Z}_{2^{64}}$ with $18$-bit fractional part. 
Our experiments are run on 3 Alibaba Cloud ecs.g7.8xlarge servers with 32 vCPU and 128GB RAM each. The CPU model is Intel Xeon(Ice Lake) Platinum 8369B CPU @ 2.70GHz. We evaluate \puma\ on Ubuntu 20.04.6 LTS with Linux kernel 5.4.0-144-generic. Our bandwidth is about 5Gbps and round trip time is about 1ms. 

\textbf{Models \& Datasets.}
We evaluate \puma\ on seven NLP models: Bert-Base, Roberta-Base, and Bert-Large~\citep{bert}; GPT2-Base, GPT2-Medium, and GPT2-Large~\citep{gpt}; and LLaMA-7B~\citep{touvron2023llama}. We measure the Bert performance for three NLP tasks over the datasets of Corpus of Linguistic Acceptability (CoLA), Recognizing Textual Entailment (RTE), Stanford Question Answering Dataset (QNLI) from GLUE benchmarks~\citep{wang2018glue}, and GPT2 performance on Wikitext-103 V1~\citep{merity2016pointer}.

\textbf{Baseline.}
We compare \puma\ to the most similar prior work \mpcformer~\citep{li2023mpcformer}. But for fair comparison, we have the following considerations:
\romannumeral1) As \mpcformer\ neither supports loading pretrained transformer models nor implements LayerNorm faithfully\footnote{ As \mpcformer~does not support loading pre-trained Transformer models, we did an experiment in plaintext Bert-Base that replaced LayerNorm with BatchNorm  as \mpcformer~did. This resulted in a significant drop in the MCC score for CoLA task from $0.616$ to $-0.020$. On the contrary, \puma~achieves an MCC score of $0.613$. }, we cannot achieve meaningful secure inference results using their framework.
Therefore, we compare our performance to that of plaintext (floating-point) to show our precision guarantee.
\romannumeral2) \mpcformer\ with \textit{Quad} approximations requires retraining the  modified models. As \puma\ does not require retraining, we compare our cost to that of \mpcformer\ without \textit{Quad} approximations. Also, we re-run \mpcformer~in our environment.

\subsection{Precision}\label{sec:accuracy}

\input{Table/accppl}

We compare our secure model 
inference performance to that of plaintext (floating-point) in Table~\ref{table:bertacc} and~\ref{tab:gpot2ppl} to show our precision guarantee.

In Table~\ref{table:bertacc}, we show the Matthews correlation/accuracy of plaintext and \puma\ on the Bert-Base, Roberta-base, and Bert-Large. We observe that the accuracy achieved by \puma~ matches the accuracy of the plaintext Flax code. Specifically, the accuracy difference does
not exceed $0.011$ over all datasets. 
Moreover, in Table~\ref{tab:gpot2ppl}, we also compare our perplexity on dataset Wikitext-103 V1 with the plaintext baseline on GPT2 models. The results are similar and the perplexity differences do not exceed $0.02$ over all models.

The above accuracy and perplexity advantages experimentally validate that our protocols are numerically precise.

\input{Table/CostsofOneInput}

\subsection{Inference Costs}\label{sec:efficiency}

We compare \puma's inference cost to that of \mpcformer. 
The costs are for processing one input sentence: \romannumeral1) For Bert models the input sentence is of length $128$. \romannumeral2) For GPT2 models the input  length is 32 and generate $1$ new word. 

On the 3 Bert models in Table~\ref{tab:costbert}, \puma\ is  $1.375\sim 1.916\times$ faster than  \mpcformer, and is $1.079\sim 1.195\times$ more communication-efficient. For the GPT2 models in Table~\ref{tab:costgpt2}, \puma\ is $2.250\sim 2.414\times$ faster than \mpcformer, and is $1.325\sim 1.884\times$ more communication-efficient. 
    
We observe that \puma's improvements increase as the model size grows, particularly for the GPT2 models. This trend is because our specialized optimizations are more effective when processing large-scale evaluations.

\subsection{Scalability}\label{sec:scala}

In this subsection, we measure the costs of evaluating \puma\ on Bert-Base and GPT2-Base models for batched inputs, varying-length inputs, and varying-length outputs (only for GPT2-Base). We also compare our costs to those of \mpcformer~to demonstrate our improvements.

\input{Table/CostInputLength}

\textbf{Input Length Evaluation.}
Table~\ref{tab:costbertinput} shows our costs on varying-length inputs, we evaluate Bert-Base on inputs of length $\{64, 128, 256\}$, and GPT2-Base on inputs of length $\{16, 32, 64\}$.
For Bert-Base, \puma\ is $1.631\sim 1.837\times$ faster, and for GPT2-Base, \puma\ is $1.744\sim 2.686\times$ faster.

\begin{figure} 
    \centering
    \includegraphics[width=250pt]{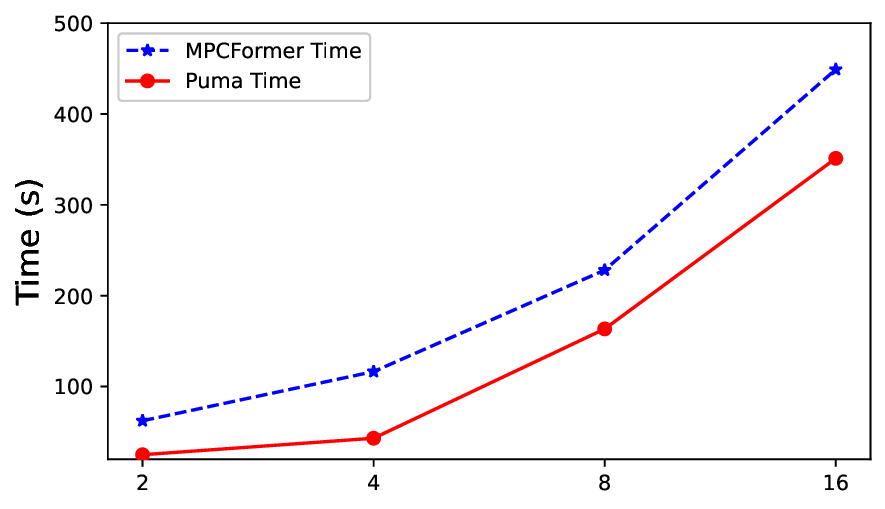}
    \vspace{-0.2cm}
    \caption{Runtime of GPT2-Base for generating different output tokens, the input length is of length $32$.} 
    \label{fig:gptwoutcosts}
\end{figure}

\textbf{Output Length Evaluation.}
Fig~\ref{fig:gptwoutcosts} presents our costs on varying-length outputs for GPT2-Base. Our improvements against \mpcformer\  range from $1.279\sim 2.700\times$.

We observe in Table~\ref{tab:costbertinput} and Fig~\ref{fig:gptwoutcosts} that for GPT-2, our efficiency gains decrease  with more input/output tokens. This is because \puma\ introduces extra one-hot embedding costs (as described in~\ref{sec:embed}). We should emphasize
again that  \puma\ is compatible with plaintext models, and could achieve a similar accuracy as plaintext models while \mpcformer\ could not.

\subsection{Evaluating LLaMA-7B in Five Minutes.}\label{sec:llama}
\input{Table/llama7b}

Our protocols are already complete for evaluating any Transformer-based models including LLaMA-7B. Unfortunately, existing serialization libraries such as Protobuf~\citep{protobuf} and FlatBuffers~\citep{van2014flatbuffers} only support data trunks with size up to 2GB, which is not sufficient for large MPC tasks. To address this problem, we propose an optimization to SecretFlow-SPU. Concretely, the system could automatically divide and serialize overly large secret-shared structures into smaller chunks when communicating or performing I/O operations.

We evaluated the large language model LLaMA-7B using \puma\ under 3 Alibaba Cloud
ecs.r7.32xlarge servers, each has 128 threads and 1TB RAM, with 20GB bandwidth, 0.1ms round-trip-time. 
As shown in Table~\ref{tab:llama7b}, \puma\ can support secure inference of LLaMA-7B with reasonable costs. For example, given an input sentence of 8 tokens, \puma\ can output one token in around $200$ seconds with communication costs of $1.794$ GB. To our knowledge, this is the first time that LLaMA-7B has been evaluated using MPC. Moreover, \puma\ can generate the same tokens exactly as plaintext LLaMA-7B, see Appendix~for an example.


%% file: Table/accppl.tex
\begin{table}
\centering
\caption{Performance on GLUE benchmark of Bert-Base, Roberta-Base, and Bert-Large on CoLA, RTE, and QNLI, Matthews correlation is reported for CoLA. Accuracy is reported for other datasets.}\label{table:bertacc}
\begin{tabular}{c|ccc|ccc|ccc}
\hline \hline
 Model & \multicolumn{3}{c|}{Bert-Base} & \multicolumn{3}{c|}{Roberta-Base} & \multicolumn{3}{c}{Bert-Large} \\ \hline
 TASK & CoLA & RTE & QNLI & CoLA & RTE & QNLI & CoLA & RTE & QNLI \\ \hline
CPU & $0.616$     & $0.700$      & $0.916$     & $0.629$ & $0.805$ & $0.920$  & $0.686$   & $0.755$ & $0.922$ \\
\puma   & $0.613$     & $0.700$     & $0.916$     & $0.618$ & $0.805$ & $0.918$ & $0.690$ & $0.747$ & $0.918$ \\ \hline \hline
\end{tabular}
\vspace{-0.2cm}
\end{table}

\begin{table}[]
\vspace{-0.2cm}
    \centering
    \caption{Perplexity of GPT2-Base, GPT2-Medium, and GPT2-Large on Wikitext-103 V1.}
    \label{tab:gpot2ppl}
    \begin{tabular}{c|c|c|c}
    \hline \hline
      Model & GPT2-Base & GPT2-Medium & GPT2-Large \\ \hline
      CPU & $16.284$ & $12.536$ & $10.142$ \\
      \puma & $16.284$ & $12.540$ & $10.161$ \\
      \hline \hline
    \end{tabular}
\vspace{-0.5cm}   
\end{table}

%% file: Table/CostsofOneInput.tex
\begin{table}[h]
    \centering
    \caption{Costs of Bert-Base, Roberta-Base, and Bert-Large for one sentence of length $128$. Time is in seconds and Communication (Comm. for short) is in GB, which is the same for the following tables.}\label{tab:costbert}
    \begin{tabular}{c|cc|cc|cc}
    \hline \hline
       Model & \multicolumn{2}{c|}{Bert-Base} & \multicolumn{2}{c|}{Roberta-Base} & \multicolumn{2}{c}{Bert-Large} \\ \hline
       Costs & Time & Comm. & Time & Comm. & Time & Comm. \\ \hline
       \mpcformer & $55.320$ & $12.089$ & $57.256$ & $12.373$ & $141.222$ & $32.577$ \\
       \puma & $33.913$ & $10.773$ & $41.641$ & $11.463$ & $73.720$ & $27.246$ \\
       \cellcolor{mygray} Improv. & \cellcolor{mygray} $1.631\times$ & \cellcolor{mygray} $1.122\times$ & \cellcolor{mygray} $1.375\times$ & \cellcolor{mygray} $1.079\times$ & \cellcolor{mygray} $1.916\times$ & \cellcolor{mygray} $1.195\times$ \\
       \hline \hline
    \end{tabular}
    \vspace{-0.2cm}
\end{table}

\begin{table}[]
    \centering
    \caption{Costs of GPT2-Base, GPT2-Medium, and GPT2-Large. The input sentence is of length $32$, all of the costs are for generating $1$ token.}\label{tab:costgpt2}
    \begin{tabular}{c|cc|cc|cc}
    \hline \hline
       Model & \multicolumn{2}{c|}{GPT2-Base} & \multicolumn{2}{c|}{GPT2-Medium} & \multicolumn{2}{c}{GPT2-Large} \\ \hline
       Costs & Time & Comm. & Time & Comm. & Time & Comm. \\ \hline
       \mpcformer & $34.889$ & $4.999$ & $73.078$ & $11.766$ & $129.095$ & $22.522$  \\
       \puma & $15.506$ & $3.774$ & $30.272$ & $7.059$ & $54.154$ & $11.952$ \\
       \cellcolor{mygray} Improv. & \cellcolor{mygray} $2.250\times$ & \cellcolor{mygray} $1.325\times$ & \cellcolor{mygray} $2.414\times$ & \cellcolor{mygray} $1.667\times$ & \cellcolor{mygray} $2.383\times$ & \cellcolor{mygray} $1.884\times$ \\
       \hline \hline
    \end{tabular}
    \vspace{-0.2cm}
\end{table}

%% file: Table/CostInputLength.tex
\begin{table}[]
    \centering
    \caption{Costs of Bert-Base and GPT2-Base for different input length (denoted as \#Input). The input lengths for Bert-Base and GPT2-Base are respectively $\{64, 128, 256\}$ and $\{16, 32, 64\}$. GPT2-Base generates $1$ token.}\label{tab:costbertinput}
    \begin{tabular}{cc|cc|cc|cc}
    \hline \hline
       \multicolumn{2}{c|}{\#Input} & \multicolumn{2}{c|}{$64 / 16$} & \multicolumn{2}{c|}{$128 / 32$} & \multicolumn{2}{c}{$256 / 64$} \\ \hline
       \multicolumn{2}{c|}{Costs} & Time & Comm. & Time & Comm. & Time & Comm. \\ \hline
       \multirow{3}{*}{Bert}& \mpcformer & $36.354$ & $5.707$ & $55.320$ & $12.089$ & $112.453$ & $29.927$ \\
       & \puma & $21.141$ & $4.881$ & $33.913$ & $10.773$ & $61.210$ & $26.004$ \\
       & \cellcolor{mygray} Improv. & \cellcolor{mygray} $1.720\times$ & \cellcolor{mygray} $1.169\times$ & \cellcolor{mygray} $1.631\times$ & \cellcolor{mygray} $1.122\times$ & \cellcolor{mygray} $1.837\times$ & \cellcolor{mygray} $1.151\times$  \\
       \hline
       \multirow{3}{*}{GPT2}& \mpcformer & $29.695$ & $4.011$ & $34.889$ & $4.999$ & $43.344$ & $7.318$  \\
       & \puma & $11.056$ & $1.875$ & $15.506$ & $3.777$ & $24.860$ & $7.821$ \\
       &\cellcolor{mygray} Improv. & \cellcolor{mygray} $2.686\times$ & \cellcolor{mygray} $2.139\times$ & \cellcolor{mygray} $2.250\times$ & \cellcolor{mygray} $1.324\times$ & \cellcolor{mygray} $1.744\times$ & \cellcolor{mygray} $0.936\times$ \\
       \hline \hline
    \end{tabular}
    \vspace{-0.3cm}
\end{table}

%% file: Table/llama7b.tex
\begin{table}[]
    \centering
    \caption{Costs of the secure inference of LLaMA-7B, \#Input denotes the length of input sentence and \#Output denotes the number of generated tokens.}\label{tab:llama7b}
    \begin{tabular}{c|cc|cc|cc}
    \hline \hline
       (\#Input, \#Output) & \multicolumn{2}{c|}{$(4,1)$} & \multicolumn{2}{c|}{$(8,1)$} & \multicolumn{2}{c}{$(8,2)$} \\ \hline
       Costs & Time & Comm. & Time & Comm. & Time & Comm. \\ \hline
       \puma & $122.004$ & $0.907$ & $200.473$ & $1.794$ & $364.527$ & $3.857$ \\
       \hline \hline
    \end{tabular}
    \vspace{-0.2cm}
\end{table}

%% file: conclusion.tex
\section{Conclusion}\label{sec:conclusion}
We propose an efficient MPC framework \puma\ for secure inference on Transformer models based on replicated secret sharing. To reduce the costs of secure inference, we approximate expensive functions with accurate polynomials and propose secure Embedding and LayerNorm protocols to support end-to-end secure inference. Although the inference cost is still quite high, we successfully make it one step closer to solving users' privacy concerns in Transformer-based DLaaS. We believe that by combining \puma\ with quantization methods and hardware accelerations in the future, secure inference of large Transformer models in seconds is no longer impossible.

%% file: appendix.tex
\appendix

\section{Details of Experimental Models}\label{models}
In this section, we present the architecture of the experimental models in brief. For more details, please refer to HuggingFace Transformers library~\citep{huggingfacetransformers}.
\begin{itemize}
    \item Bert-Base: Bert-Base is the base version of the Bert model and consists of $12$ Transformer encoder layers, $768$ hidden size, and $12$ heads. It has $110$ million parameters and is trained on a large corpus of unlabeled text data.
    
    \item Roberta-Base: Similar to Bert-base, Roberta-base is a base version of the Roberta model. It comprises $12$ Transformer layers, $768$ hidden size, and $12$ heads. It has around 125 million parameters.

    \item Bert-Large: Bert-Large is an extended version of Bert-base with $24$ Transformer encoder layers, $1024$ hidden size, and $16$ heads. It has approximately $340$ million parameters, making it more powerful and capable of capturing complex language patterns.

    \item GPT2-Base: GPT2-Base is the base version of the Gpt2 model and consists of $12$ Transformer decoder layers, $768$ hidden size, and $12$ heads. It has $117$ million parameters and is trained on a large corpus of text data. GPT2-Base is mainly used for tasks involving text generation and language understanding.

    \item GPT2-Medium: GPT2-Medium comprises $24$ Transformer decoder layers, $1024$ hidden size, and $16$ heads. And it has approximately $345$ million parameters. 

    \item GPT2-Large: GPT2-Large is the largest variant of the GPT2 model, featuring $36$ Transformer decoder layers, $1280$ hidden size, and $16$ heads.
    It has approximately $774$ million parameters.
\end{itemize}


\section{\puma\ for LLaMA-7B}
Unlike GPT-2 and Bert, LLaMA uses SiLU instead of GeLU, we can approximate SiLU using similar piece-wise low-degree polynomials with different coefficients. The full polynomials could be found in $flax\_llama7b.py$ . 

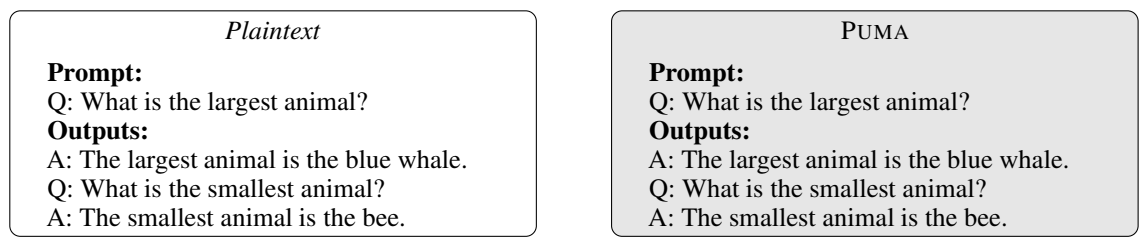
\begin{figure}
  \centering
  \begin{tikzpicture}

    \node[draw, rectangle, rounded corners, minimum width=7cm, minimum height=3cm] (left) at (0,0) {
     \begin{minipage}{6cm}
          \begin{center}
              \textit{Plaintext}
          \end{center}
          \textbf{Prompt:}\\
          \ \ Q: What is the largest animal?\\
          \textbf{Outputs:}\\
          \ \ A: The largest animal is the blue whale.\\
          \ \ Q: What is the smallest animal?\\
          \ \ A: The smallest animal is the bee.
      \end{minipage}
    };
    
    \node[draw, rectangle, rounded corners, minimum width=7cm, minimum height=3cm, fill=gray!20] (right) at (8,0) {
      \begin{minipage}{6cm}
        \begin{center}
            \textit{\puma}
        \end{center}
          \textbf{Prompt:}\\
          \ \ Q: What is the largest animal?\\
          \textbf{Outputs:}\\
          \ \ A: The largest animal is the blue whale.\\
          \ \ Q: What is the smallest animal?\\
          \ \ A: The smallest animal is the bee.
      \end{minipage}
    };

  \end{tikzpicture}
  \caption{Outputs of LLaMA-7B in plaintext and \puma.}\label{fig:llamaout}
\end{figure}

In Figure~\ref{fig:llamaout}, we show the output tokens of LLamA-7B (with fixed randomness) given the prompt: \textit{Q: What is the largest animal?}  It can be seen that our \puma\ outputs the same tokens as LLaMA-7B does in plaintext for generating more than 20 tokens. 

